\documentclass[twocolumn,showpacs,preprintnumbers]{revtex4}
\usepackage{epsfig,amsmath,amssymb}
\usepackage{graphicx}
\usepackage{dcolumn}
\usepackage{bm}
\usepackage{subfigure}
\usepackage{color}
\newcommand{\be}{\begin{equation}}    
\newcommand{\ee}{\end{equation}}
\newcommand{\beq}{\begin{eqnarray}}
\newcommand{\eeq}{\end{eqnarray}}
\newcommand{\beqn}{\begin{eqnarray*}}
\newcommand{\eeqn}{\end{eqnarray*}}
\newcommand{\sun}{\odot}

\definecolor{magenta}{cmyk}{0.1,0.8,0,0.1} 

\begin{document}

\title{Imprint of the merger and ring-down on the 
gravitational wave background from  black hole binaries coalescence}
\author{S. Marassi$^{1}$\thanks{E-mail:stefania.marassi@roma1.infn.it}, R. Schneider$^{2}$, G. Corvino$^{1}$, V. Ferrari$^{1}$, S. Portegies Zwart$^{3}$}
\affiliation{ $^{1}$ Dipartimento di Fisica `G. Marconi', Sapienza Universit\`a di Roma and Sezione INFN Roma1, Piazzale Aldo Moro 5, Roma, 00185, Italy\\
$^{2}$ INAF, Osservatorio Astronomico di Roma, via di Frascati 33, 00040 Monteporzio Catone, Italy\\
$^{3}$ Leiden Observatory, Leiden University, P.O. Box 9513, 2300 RA Leiden, The Netherlands}

\date{\today}
\label{firstpage}

\begin{abstract}

We compute the gravitational wave background (GWB) generated by a
cosmological population of (BH-BH) binaries using hybrid waveforms
recently produced by numerical simulations of (BH-BH) coalescence, which
include the inspiral, merger and ring-down contributions.  A large
sample of binary systems is simulated using the
population synthesis code SeBa, and  we extract fundamental
statistical information on (BH-BH) physical parameters (primary and
secondary BH masses, orbital separations and eccentricities, formation
and merger timescales). We then derive the binary birth and merger rates
using the theoretical  cosmic star formation history obtained from a numerical study
which reproduces the available observational data at redshifts $z < 8$. 
We evaluate the contributions of the inspiral, merger and ring-down signals
to the GWB, and discuss how these depend on the parameters which critically
affect the number of coalescing (BH-BH) systems.
  
We find that Advanced LIGO/Virgo have a chance to detect the GWB signal
from the inspiral phase with a $(S/N)=10$ only for the most optimistic
model, which predicts the highest local merger rate of
$0.85$~Mpc$^{-3}$~Myr$^{-1}$. 
Third generation detectors, such as ET, could reveal the GWB from the 
inspiral phase predicted by any of the considered models.  
In addition, ET could sample the merger phase of the
evolution at least for models which predict local merger rates
between $[0.053 - 0.85]$~Mpc$^{-3}$~Myr$^{-1}$, which are more than
a factor 2 lower the the upper limit inferred from the analysis of the LIGO S5 run
\cite{Abadieetal2011}.  

The frequency dependence and amplitude of the
GWB generated during the coalescence is very sensitive to the adopted
core mass threshold for BH formation. This opens up the possibility to
better understand the final stages of the evolution of massive stellar
binaries using observational constraints on the associated gravitational
wave emission.  

\end{abstract}  \maketitle

\section{Introduction}
Double black hole binaries are among the most promising sources of gravitational 
radiation for the ground-based detectors  LIGO/Virgo -- in their present and advanced
configurations which plan to increase the sensitivity by a factor 10
(http://www.ligo.caltech.edu/, http://www.ego-gw.it/) --
and for the future interferometers,
the space detector LISA (http://sci.esa.int/lisa) and the Einstein
Telescope (ET, www.et-gw.eu).  In addition to the emission from single
resolved systems,  massive compact binaries 
generate stochastic backgrounds of gravitational waves (GWB), as
extensively discussed in the literature 
\cite{SchFerMat2001,IgnKurPos2001,NelYunPor2001a,RegdeF2006,RegCha2007,BelBenBul2010,ZhuHowReg2011,Reg2011}.
If detected, these backgrounds would provide information on
the cosmic star formation history, on the evolution of compact stars in binary
systems, and on cosmological parameters
\cite{CutHol2009,TayGaiMan2011}; even if these signals were not detected, more 
stringent upper limits would rule out the most optimistic theoretical models.
Astrophysical backgrounds act as a confusion-limited foreground noise for signals
generated in the early Universe (primordial GWBs); therefore, the spectral properties
which characterize the background generated by
 various families of compact binaries need to be accurately
modeled, and specific techniques need to be envisaged in order to disentangle 
their contribution from the instrumental noise \cite{AllRom1999,Mag2000,RegMan2008}.

The aim of the present work is to provide updated estimates of the GWB generated 
by a cosmological population of (BH-BH) binaries, including the contribution due 
to the merging of the two bodies, and to the ring-down of the final black hole. 
The full signal is modeled using the waveforms derived in 
\cite{AjiBabChe2008}; they combine the inspiral part of the signal 
(obtained with the standard Post-Newtonian description) and the ring-down 
oscillations of the final black hole, with the signal emitted during the 
merging phase, whose description has been made possible by recent progresses 
in numerical relativity. We will show that since a significant amount of 
energy is radiated during the merger and ring-down, these phases give a 
significant contribution to the produced background. 

A recent estimate of GWB from coalescing black hole binaries using the
same waveforms has been presented in \cite{ZhuHowReg2011}. 
This study adopts average quantities for the single source emission and
a fixed distribution for the delay between the birth of a binary and
its merging (merger timescale); furthermore, they normalize the (BH-BH) merger rate to a
``local value'' inferred from models of Galactic binary populations. 

Here we follow a different approach. We use the
population synthesis code SeBa \cite{YunLasNel2006}
(http://www.sns.ias.edu/$\sim${\tt starlab/}) to simulate the properties
of a large sample of double black hole binaries.
This enables us to investigate the statistical distributions of masses
and merger timescales in a self-consistent way.
Following \cite{MarSchFer2009,MarCioSch2011}, we adopt a theoretical model for the
cosmic star formation history at redshifts $z < 20$ taken from
the numerical simulation of \cite{TorFerSch2007}, which reproduces
the observational data available at $z < 8$ \cite{BouIllFra2010}.
With these inputs, we compute the redshift evolution of the (BH-BH)
birth and merger rates and the cumulative gravitational wave
signal produced throughout the Universe.
We evidenciate the contributions of the inspiral, merger and ringdown
phases of the coalescence process to the total background, and discuss
their detectability.

Finally, we run a set of independent SeBa simulations to single out 
the parameters which critically affect the number of 
coalescing (BH-BH) systems, the GWB and its detectability; 
we find these to be  the common envelope parameter, the 
core mass threshold for BH formation and the kick velocity distribution.

As first noted in \cite{PorMcM2000}, close black hole binaries form 
efficiently through dynamical interactions in dense globular clusters 
or in nuclear star clusters \cite{LeaShaRas2007,SadBelBul2008,MilLau2009}. 
However, in the present analysis we do not consider this additional formation
channel, since it requires a different modeling which is not included in the 
current version of SeBa.

The plan of the paper is the following.    
In section \ref{sec:seba} we briefly describe the SeBa code and the
underlying physical assumptions  which
we use to simulate the sample of (BH-BH) binaries.
A statistical analysis of the resulting population is
presented in section \ref{sec:stat}.
In section \ref{sec:gwsignal} we outline the main features of the  
waveforms we adopt to describe the gravitational emission of (BH-BH)
binaries and in section \ref{sec:rates} we compute, starting from the
cosmic star formation history, the evolution of the
birth and merger rates of (BH-BH) binaries.
In section \ref{sec:gwback}, we present the resulting density 
parameter of the GWB, $\Omega_{\rm GW}$, we discuss its detectability 
by second and third generations interferometric detectors, 
and its dependence on  key physical parameters. 
Finally, in section \ref{sec:conclusions} we draw our conclusions.

\section{SeBa population synthesis code}
\label{sec:seba}
To compute the statistical properties of a black hole binary population
we use the latest release of the population synthesis code SeBa \cite{YunLasNel2006}, 
which is based on previous versions described in 
\cite{PorVer1996,PorYun1998,NelYunPor2001a,NelYunPor2001b}.

In SeBa, the evolution of binary systems is
modeled by taking into account the relevant physics,
which include stellar composition, stellar wind, mass transfer and accretion, 
gravitational radiation, magnetic braking, common envelope phase and supernovae 
(for more details see the original papers).  
The present version of the code uses updated stellar and binary physics, 
including results from supernova simulations \cite{FryKal2001}. 
In particular, new features about the evolution of massive stars, Wolf-Rayet
stars, stellar wind mass loss rate, common envelope phase, fallback
prescription and supernova kicks are discussed in detail in \cite{YunLasNel2006}.
The predictions of the code have been tested against observational data
on double neutron star binaries \cite{PorYun1998}, BH-candidate short-period binaries 
\cite{YunLasNel2006}, Type II and Ib/c SNe \cite{NelYunPor2001a,NelYunPor2001b}, and 
more recently, SN Type Ia \cite{TooNelPor2010}. 
Here we sketch a summary of the 
assumptions that are relevant for the evolution of (BH-BH) binaries.

For the common envelope evolution (CE), we use the standard prescription
described in \cite{Web1984,deKHeuPyl1987,NelYunPor2001a}, with the
efficiency and the structure parameters $\alpha_{\rm CE}$ and $\lambda$
combined into a single quantity, $\alpha_{\rm CE} \times \lambda$.  The
treatment of the common envelope evolution is still under debate; at
present, a strict criterion to define the binding energy of the stellar
envelope is still lacking and it is unclear whether other sources of
energy, beyond gravitational energy, contribute to unbind the common
envelope
(\cite{PodRapHan2003,TauDew2001,NelTou2005,KieHur2006,YunLas2008,DelTaa2010}
and references therein). As a result, $\alpha_{\rm CE} \times \lambda$
is a free parameter of population synthesis models.
For our reference model A (see Section \ref{sec:stat}), we fix
$\alpha_{\rm CE} \times \lambda = 2$, chosen so as to
reproduce the available observational constraints \cite{PosYun2006},
such as the (NS-NS) merger rate inferred from double pulsar
observations \cite{Burgayetal2003,Kaletal2004}.

As suggested in \cite{JonNel2004,WilHenLev2005,GuaColPor2005}, black holes 
receive a small asymmetric kick at birth: in our reference simulation
(model A) this kick 
is taken from a Paczy\'nsky velocity distribution \cite{Pac1990,Har1997},  
isotropic in space and scaled down with the ratio of black hole to neutron 
star masses (see Table \ref{ModelA}).

Black hole formation is treated in the code assuming that a constant fraction
of the supernova explosion energy is used to unbind the stellar envelope \cite{FryKal2001}; 
we choose $f=0.4$. However, while in \cite{FryKal2001} the explosion energy is assumed to be a
function of the pre-supernova mass, we keep it fixed at 10$^{50}$ erg. This value is within 
the expected range, but favours the formation of rather massive black-holes 
(up to 15 $M_\odot$, see also \cite{YunLasNel2006}).

\begin{table*}
\begin{minipage}{176mm}
\centering
\begin{center}
\caption{Zero-age  parameters for the reference Model A (see text).}
\begin{tabular}{|c|c|c|c|}
\hline
\multicolumn{4}{|c|}{Model A}\\
\hline
 Parameter & Symbol& Value & Note\\
\hline
Mass of primary star & $M_{\rm prim}$ &[ 8-100] $M_{\sun}$& Salpeter IMF (-2.35)\\
\hline
Mass of the secondary star &$M_{\rm sec}$ & $M_{\rm sec} = q M_{\rm prim}$ & the distribution matches the q-distribution\\
\hline
Mass ratio & $q$ & [0 - 1] & flat distribution \\
\hline
Initial semi-major axis & sma & $0.1-10^{6}R_\sun$& flat distribution in log sma \\
\hline
Eccentricity & $e$ & [0 - 1] & thermal equilibrium distribution\\
\hline
CE parameter &$\alpha_{CE}\lambda$&2& structure parameter\\
\hline
Kick distribution for NS &$u=v/\sigma$& $\sigma=300$~km s$^{-1}$& Paczy\'nski distribution for $v$\\
\hline
Kick distribution for BH &$v_{\rm BH}$ & -  & same as NS but scaled down: $v_{\rm BH} = v(M_{\rm NS}/M_{\rm BH})$ \\
\hline
Core mass threshold  & $m_{\rm thre, BH}$ & $10 M_{\odot}$ & \cite{YunLasNel2006} \\
\hline
\end{tabular}
\label{ModelA}
\end{center}
\end{minipage}
\end{table*}

Regarding the evolution of massive ($  15~M_\odot\leq M_{\rm i}\leq 85
M_{\odot} $) stars, 
SeBa assumes that stellar wind mass-loss rate increases in time;
in their total lifetime, stars lose an amount of matter which is a function
of their initial mass, $0.01 M_{\rm i}^2$;
the hydrogen envelope  is lost when the stars are still on the main
sequence.
For higher masses ($M_{\rm i} > 85 M_{\odot}$) the mass loss rate 
is very uncertain: we assume that, during the main sequence lifetime, 
these stars lose 43~$M_{\sun}$. Due to the paucity of these massive stars, 
this crude assumption has negligible consequences on the estimate of the GWB.

When the star loses its hydrogen envelope on the main sequence, 
stellar mass loss prescription for Wolf-Rayet stars is adopted.
More details on the treatment of massive stars in the code
are described in \cite{YunLasNel2006} (see in particular section 2.3 and Fig.1).
For Wolf-Rayet stars the mass loss rate by \cite{NelHeu2001} is adopted, which 
is based on the compilation of mass-loss rates inferred from
observations of  Wolf-Rayet stars \cite{NugLam2000}.

We initialise $N = 10^{6}$ ``zero-age'' binaries (ZAMS). The zero-age
parameters of the simulated population are given in Table \ref{ModelA} 
for our reference model A. These are randomly selected from a set of 
independent distribution functions. In particular, the initial primary 
mass $M_{\rm prim}$ is taken from a Salpeter Initial Mass Function (IMF), 
$\Phi(M)\propto M^{-(1+x)}$ with $x=1.35$, between [8-100]$M_\sun$; the 
initial secondary mass, $M_{\rm sec}$, is selected from a flat distribution for 
the mass ratio $q = M_{\rm sec}/M_{\rm prim}$. The semi-major axis (sma) 
distribution is flat in log(sma) \cite{Abt1983} ranging from 
0.1 $R_\sun$ (Roche lobe contact) up to $10^{6}R_\sun$. We assume a 
thermal eccentricity distribution $\Phi(e)=2e$ in the range [0-1] 
\cite{DuqMay1991}. Kicks follow a Paczy\'nski velocity distribution 
with a dispersion of $\sigma = 300$~km s$^{-1}$ \cite{Pac1990}.

We have explored a wide parameter space to investigate how different
choices  affect the resulting GWB. This has enabled us to identify the 
physical parameters which have the largest impact on the GWB,
and the corresponding models will be presented
in section \ref{sec:par}. 

Finally, we note that the version of SeBa that we use assumes an
initial solar  metallicity for all stars.

\section[]{Statistical properties of black hole binaries population}
\label{sec:stat}
In this section, we present the statistical properties of (BH-BH) 
binaries and their progenitors for our reference model A 
(see Table \ref{ModelA}). A great uncertainty in the modeling of black 
hole binaries is due to our poor knowledge of the BH mass distribution 
and its relation to the initial distribution of progenitor masses. The 
upper mass limit of isolated stars for black hole formation depends 
predominantly on wind mass loss; recent studies  constrain 
the mass range of black hole progenitors to $M \sim [20 - 60]M_{\sun}$ 
\cite{PorVerErg1997,FryKal2001,PosYun2006,BelDomBul2010}. Clearly, a 
massive progenitor in a binary system might follow a different evolutionary 
path which would affect the mass of the nascent black hole 
\cite{FryKal2001,NelHeu2001,FryHegLan2002,ZhaWooHeg2008}.

\begin{figure}
\includegraphics[width=8.5cm,angle=360]{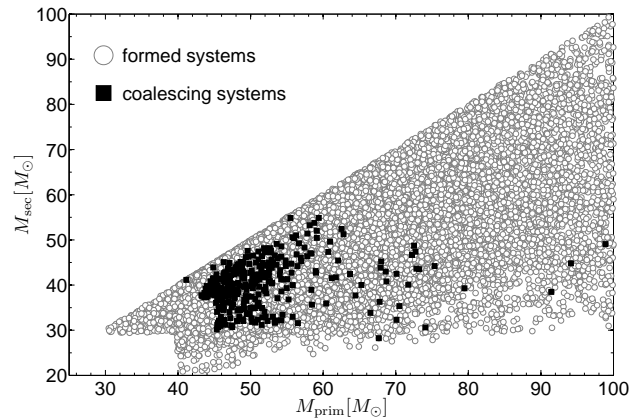}
\caption{Mass of the secondary stellar progenitors as a function of the
corresponding primary mass. Empty grey circles indicate stellar
binaries which lead to (BH-BH) systems; black squares show the subsample
of these progenitors which form (BH-BH) binaries with merger times
smaller than the Hubble time (see text).}
\label{fig:bhprogmass}
\end{figure}

Figure \ref{fig:bhprogmass} shows the mass range of primary and
secondary progenitors of (BH-BH) systems (empty grey circles). Black
squares show the sub-sample of progenitors which generate (BH-BH)
binaries with merger times smaller than the Hubble time ($t_{\rm H} \sim
13.4$~Gyr for our adopted cosmological model \footnote{We adopt a
$\Lambda$CDM cosmological model with parameters $\Omega_{\rm M} = 0.26$,
$\Omega_{\rm \Lambda} = 0.74$, $H_0=73$~km/Mpc/s, $\Omega_{\rm B}
=0.041$.  }).  Primary (secondary) progenitors have masses in the range
$30 M_{\sun} \le M_{\rm prim} \le 100 M_{\sun}$ ($20 M_{\sun} \le M_{\rm
sec} \le 100 M_{\sun}$).  Merging systems come from a narrower dynamical
range, with 90\% (96 \%) having primary (secondary) progenitor mass $30
M_{\sun} \le M_{\rm prim} \le 60 M_{\sun}$ ($20 M_{\sun} \le M_{\rm sec}
\le 50 M_{\sun}$).

\begin{figure}
\centering
\includegraphics[width=8.5cm,angle=360]{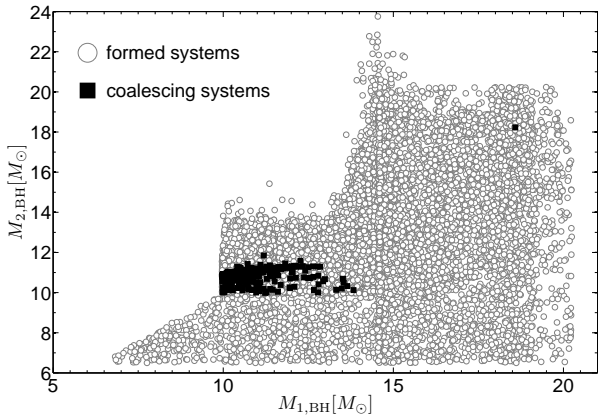}
\caption{Same as Fig.\ref{fig:bhprogmass} but for black hole
primary and secondary masses. The various regions in the parameter space
which are over- or under-populated 
originate from the
decision making process in SeBa, and can be related to the chain of
events in the
population synthesis.}
\label{fig:bhmass}
\end{figure}
Observations of BH candidates in binary systems suggest a range of
masses in the $[4 - 17] M_{\sun}$ interval
(\cite{PosYun2006,ManShau2010,Ozeletal2010} and references therein),
in broad agreement with theoretical simulations, which indicate a
range  of black hole masses
up to $[10 - 15] M_{\sun}$ \cite{FryKal2001,ZhaWooHeg2008}.
In Fig.~\ref{fig:bhmass} we plot
the  primary and secondary black hole masses, $M_{1,\rm BH}$ and
$M_{2, \rm BH}$  (black squares) for model A.
Primary and secondary black holes have comparable masses, ranging
between $\sim 6 M_{\sun}$ up to $\sim 20 M_{\sun}$; these limits are
consistent with the observational and theoretical estimates quoted
above. 

Merging  pairs are concentrated in a narrow range of masses,
with the primary  BH mass  in the range $[10 - 15] M_{\sun}$ and the secondary 
in the range $[10 - 11] M_{\sun}$. This corresponds to 
a BH mass ratio, $q_{\rm BH}$, in the range [1 - 1.1] (see also
Fig.~\ref{fig:ecc}), where $q_{\rm BH}$ is defined as the ratio of the heaviest to the 
lightest BH mass.

\begin{figure}[ht]       
\includegraphics[width=8.5cm,angle=360]{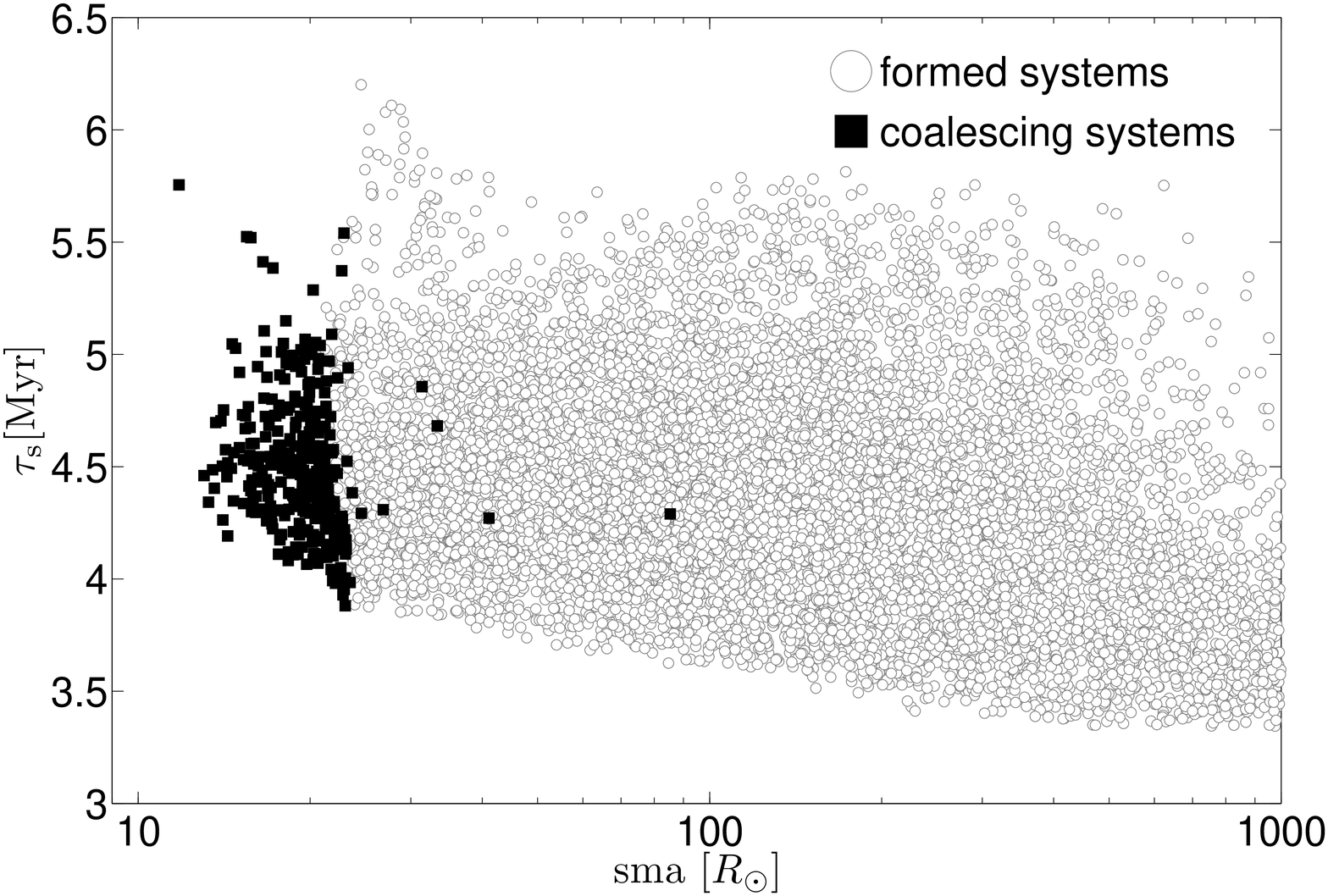}
\includegraphics[width=8.5cm,angle=360]{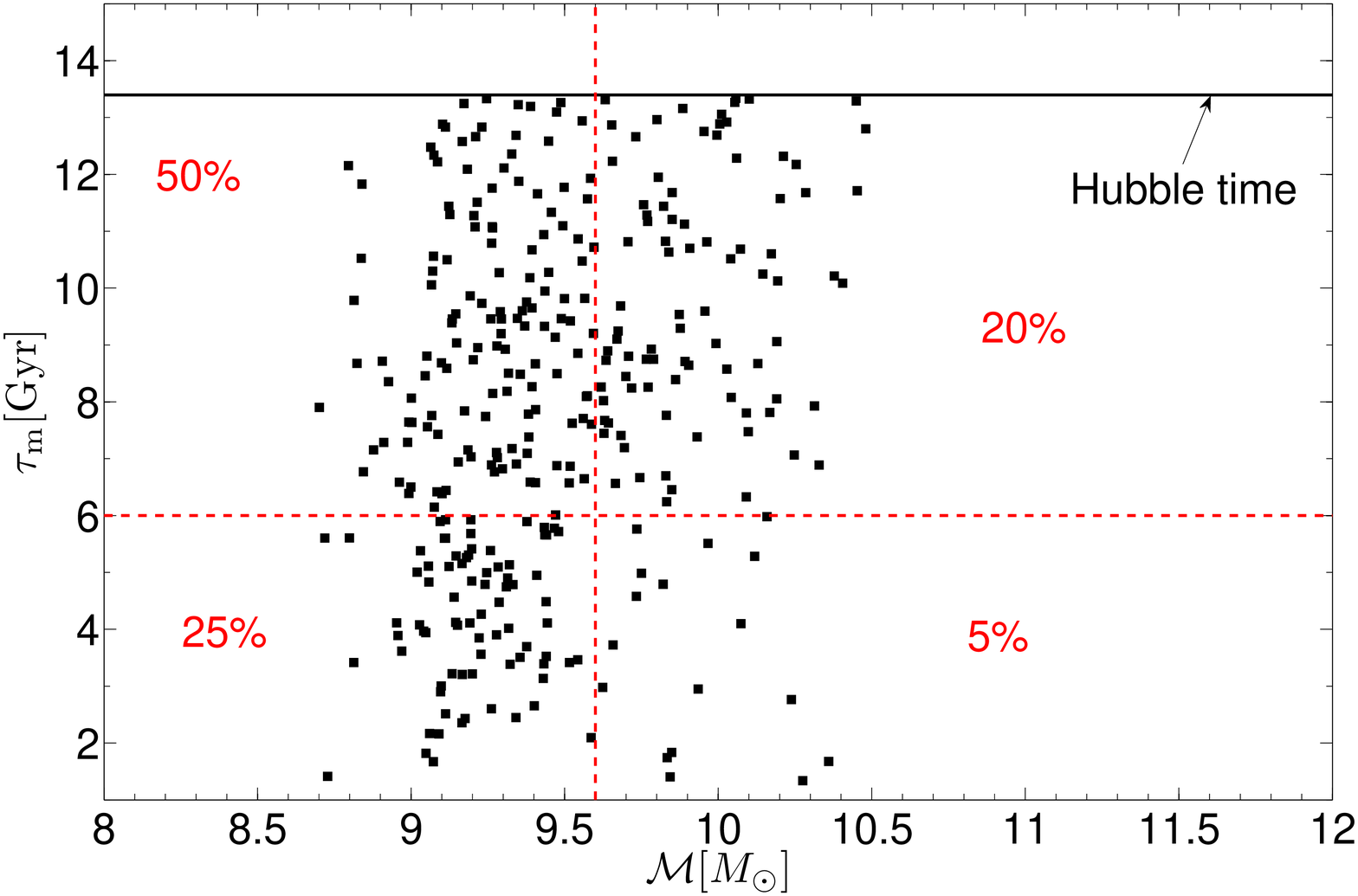}
\caption{ {\it Upper panel}: formation timescales for compact black
hole binaries as a function of the corresponding semi-major axis.
For clarity, only systems with sma $< 1600 R_{\sun}$ are shown (empty
grey circles).
Black squares are (BH-BH) pairs with merger times
smaller than the Hubble time (see text). {\it Lower panel}: merger
timescales as a function of the chirp mass of merging systems. 
The dashed horizontal and vertical
lines indicate threshold values, $\tau_{\rm m, thre}$ and ${\cal M}_{\rm thre}$, 
so that 70\% of the systems have $\tau_{\rm m} > \tau_{\rm m, thre}$ and 
75\% have ${\cal M} < {\cal M}_{\rm thre}$.}
\label{fig:times}
\end{figure}

\begin{figure}
\centering
\includegraphics[width=8.5cm,angle=360]{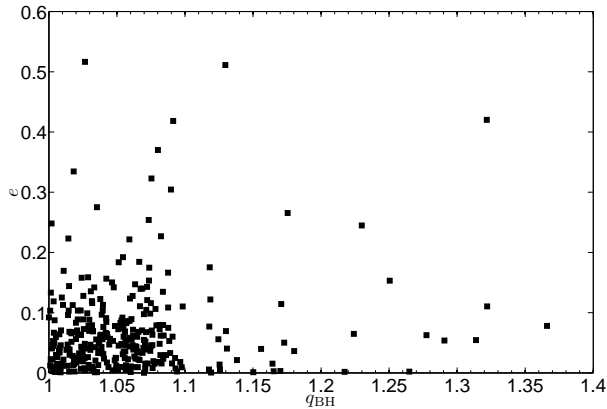}
\caption{Eccentricity as a function of the black hole mass ratio for
merging systems.}
\label{fig:ecc}
\end{figure}

In the upper panel of Fig.~\ref{fig:times}, we show  the
time interval from the formation of the ZAMS binary system to the
formation of the compact black hole binary, $\tau_{\rm s}$; we plot this quantity as a
function of the semi-major axis, sma, showing only those systems with
sma $< 1600~R_{\sun}$, which represent $\sim  60$\% of the total.  The
formation time of compact binaries is very small, ranging between $\sim
3.5$ to $\sim 6$~Myr.  Black squares indicate merging black hole
binaries, which are those that, at the time of their formation, have
semi-major axis smaller than $20~R_{\sun}$, comparable to what found
in \cite{GriLipPos2001} (see their Fig.2).

It is important to note that only 2\% of (BH-BH) pairs is able to reach the final
coalescence. Figure~\ref{fig:times} shows that the majority of (BH-BH) systems 
is characterized by large orbital separations (and periods). In fact, progenitors
with masses $\geq 40 M_{\odot}$ experience large mass loss rates which remove mass
from the binary increasing the orbital separation. 

Once compact degenerate binaries are formed, the emission of
gravitational radiation is the only physical process driving the change
in orbital parameters.  In the lower panel of Fig.~\ref{fig:times}, we
show the merger time as a function of the chirp mass, defined as 
\[
{\mathcal M} = \mu^{3/5} M^{2/5},
\]
\noindent
where $\mu = M_{1,\rm BH} M_{2, \rm BH}/M$ and $M=M_{1, \rm BH}+M_{2, \rm BH}$ are
the reduced and total mass, respectively. 
It is clear from the figure that all merging pairs have
relatively long merger times,
$\tau_{\rm m} > 1$~Gyr and 70\% have $\tau_{\rm m} > 6$~Gyr (horizontal dashed
line); the chirp masses lie in the interval $[8.5 - 10.5]M_{\sun}$ and
 75\% of the systems have ${\cal M} < 9.6 M_{\odot}$ (vertical dashed line). 
Although the plot does not evidenciate any clear 
correlation between $\tau_{\rm m}$ and ${\cal M}$, half of the systems lie in
the upper left region and only 5\% in the bottom right.
       
Finally, in Fig.~\ref{fig:ecc} we show the distribution of orbital
eccentricity as a function of the black hole mass ratio for merging
pairs at the time of formation of the compact binary systems.  The
inspiral part of the adopted waveform depends on the BH masses 
and it is strictly applicable
only to (BH-BH) binaries in quasi-circular orbit. From  Fig.~\ref{fig:ecc} we
see that only 10\% of the systems have $e > 0.15$.  Furthermore,  using
the relation between the orbital separation and the eccentricity due to
gravitational wave emission \cite{Pet1964,PetMat1963}, we find that by
the time (BH-BH) binaries enter the Advanced LIGO and ET bands (10~Hz
and 1~Hz, respectively), the eccentricities are smaller than $10^{-4}$
for all systems, and consequently the assumption of quasi-circular orbits is
well justified.

\section[]{Black hole binaries as GW sources}
\label{sec:gwsignal}
To evaluate the background produced by a cosmological population of coalescing
(BH-BH) binaries, we use the family of waveforms obtained 
in \cite{AjiBabChe2008}, which model the inspiral, merger and ring-down phases 
for the coalescence of non spinning black holes.
These waveforms refer to the leading harmonic of the 
gravitational signal ($\ell=2, m=\pm2$), which is the dominant contribution 
for low mass ratios ($q_{\rm BH} \sim 1$). In the frequency domain, the signal 
has the form
\be
h(f)= A_{\rm eff}(f) e^{i \Psi_{\rm eff}(f)},
\label{signal}
\ee
where $f$ is the emission frequency. The wave amplitude is given by
\be
\label{ampl}
A_{\rm eff}(f) = C
 \left\{
\begin{array}{l l}
\left(\frac{f}{f_{merg}}\right)^{-7/6}  & \quad \hbox{if}\quad f < f_{merg}\\
\left(\frac{f}{f_{merg}}\right)^{-2/3}  & \quad \hbox{if}\quad f_{merg} < f <
f_{ring}\\
w \cal{L}           & \quad\hbox{if}\quad f_{ring} < f < f_{cut}\\
\end{array} \right. ,
\ee
where the constants $(f_{merg},f_{ring},f_{cut})$, which identify the frequency regions
where the emitting system is  inspiralling, merging and oscillating, are
\beq
f_{merg}&=& \frac{a_0 \eta^2 +b_0 \eta +c_0}{\pi M}, \\\nonumber
f_{ring}&=& \frac{a_1 \eta^2 +b_1 \eta +c_1}{\pi M}, \\\nonumber
f_{cut}&=& \frac{a_3 \eta^2 +b_3 \eta +c_3}{\pi M}. 
\label{f_vari}
\eeq
In these expressions $\eta = M_{1,\rm BH} M_{2, \rm BH}/M^2$ is the symmetric mass
ratio, and the coefficients $a_k,b_k,$ and $c_k$ (with $k = 0,1,2,3$) are given in Table \ref{table_ampl}.
\begin{table}
\begin{center}
\begin{tabular}{c|c|c|c}
\hline
\hline
$k$  & $a_k$  & $b_k$  & $c_k$ \\
\hline
\hline
0&$6.6389\times 10^{-1}$&$-1.0321\times 10^{-1}$&$1.0979\times 10^{-1}$\\
1&$1.3278$&$-2.0642\times 10^{-1}$&$2.1957\times 10^{-1}$\\
2&$1.1383$&$-1.7700\times 10^{-1}$&$4.6834\times 10^{-2}$\\
3&$1.7086$&$-2.6592\times 10^{-1}$&$2.8236\times 10^{-1}$\\
\hline
\end{tabular}
\end{center}
\caption{Values of the constants which appear in the wave amplitude (eq.~\ref{ampl})
taken from \cite{AjiBabChe2008}.}
\label{table_ampl}
\end{table}
The constants $C$ and $w$, and the function $\cal{L}$, which characterize the wave amplitude, are
\beq
C&=&\frac{M^{5/6}}{d~\pi^{2/3}~f_{merg}^{7/6}}\left(\frac{5 \eta}{24}\right)^{1/2},
\\\nonumber
w&=&\frac{\pi \sigma}{2}\left(\frac{f_{ring}}{f_{merg}}\right)^{-2/3},
\label{w}
\\\nonumber
\cal{L}&=& \left(\frac{1}{2\pi}\right) \frac{\sigma}{(f-f_{ring})^2+\sigma^2/4},
\label{l}
\eeq
where
$\sigma= (a_2 \eta^2 +b_2 \eta +c_2)/\pi M$.
We do not need to compute the phase of the signal since the single source GW
spectrum depends on the squared GW amplitude (see eq.~\ref{eq:singlesource}).
The model assumes optimal orientation 
of the detector with respect to the emitting source.

\begin{figure}
\includegraphics[width=9.0cm,angle=360]{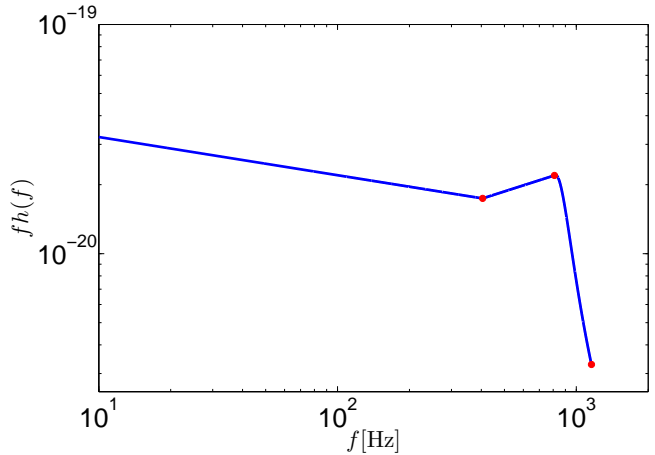}
\caption{(color online) Single source signal as a function of frequency for a (BH-BH) 
binary with total mass $M=20 M_\sun$,  symmetric mass ratio $\eta =
0.25$, located at a distance of 10~Mpc.} 
\label{fig:waveform}
\end{figure}

In Fig.~\ref{fig:waveform} we plot the dimensionless amplitude $f h(f)$
as a function of frequency, for a  system with total mass  $M = 20
M_\sun$, symmetric mass ratio  $\eta = 0.25$, located 
at a distance of 10~Mpc. 
The frequency limits, which identify the three regimes (inspiral, merger
and ring-down), are: $f_{merg} = 405$~Hz, $f_{ring}=810$~Hz, and
$f_{cut} = 1042$~Hz (points on the curve). 
These frequencies are inversely
proportional to the total mass of the binary. Hence, for the most
massive binaries in the simulated sample, with $M = 40 M_\sun$ (see
Fig.~\ref{fig:bhmass}), $f_{merg}$, $f_{ring}$, and $f_{cut}$ are a
factor of 2 smaller than the values plotted in the figure.
 
\section[]{From star formation to binary formation rate}
\label{sec:rates}
In this section we derive the evolution of the birth rate of binary
systems from the comoving star formation rate density as a function of
redshift,  inferred
from the simulations in \cite{TorFerSch2007}.  These cosmological
simulations are characterized by an improved treatment of metal
enrichment and the stellar IMF is assigned depending on the gas
metallicity. In particular, Population II/I stars form in the mass range
$[0.1 - 100] M_\sun$ according to a Salpeter IMF in regions which have
been already polluted by the first metals and dust grains to a
metallicity $Z > Z_{\rm cr} = [10^{-6} - 10^{-4}] Z_{\sun}$
\cite{SchFerNat2002,SchFerSal2003,OmuTsuSch2005}.  Below this threshold,
gas cooling is inefficient and the star formation process favors the
formation of very massive (Population III) stars, characterized by a
top-heavy IMF (for more details on the numerical scheme we refer to
\cite{TorFerSch2007}). In this work, we are only interested to
Population II/I stellar progenitors of black holes. 

Starting from the star formation rate density at a given $z$, 
$\dot{\rho}_{\star}(z)$ (expressed in units of $M_\sun$ yr$^{-1}$Mpc$^{-3}$), 
we derive the binary birth rate per comoving volume (expressed in units of yr$^{-1}$ Mpc$^{-3}$) as,
\be
\dot{R}_{\rm bin}(z)=\frac{dR}{dtdV}(z)=\frac{\dot{\rho}_{\star}(z)}{\langle m_\star \rangle}\times \frac{f_{\rm bin}}{2}\times f_{\rm sim},
\ee
where $f_{\rm bin}$ is the binarity fraction (which we take to be 1), $\langle m_\star \rangle$ 
is the average stellar mass and $f_{\rm sim}$ is the fraction of binaries simulated by 
the population synthesis code SeBa. The latter quantity accounts for the fact 
that, while in the original simulation of \cite{TorFerSch2007} stars are 
assumed to have masses in the  range [0.1-100]$M_\sun$, 
in SeBa we initialize only binary systems with primary mass in the range [8-100]$M_\sun$, 
in order to increase the statistics on double (BH-BH) binaries. 
Thus, the fraction of simulated systems is,
\be
f_{\rm sim}=\frac{\int^{100}_{8}dM_{\rm prim} \Phi(M_{\rm prim})}{\int^{100}_{0.1}dM_{\rm prim} \Phi(M_{\rm prim})},
\ee
\noindent 
and the average stellar mass is,
\be
\langle m_{\star} \rangle=\frac{\int^{100}_{0.1}dM_{\rm prim} M_{\rm prim} \Phi(M_{\rm prim})}{\int^{100}_{0.1}dM_{\rm prim} \Phi(M_{\rm prim})}.
\ee
\noindent 
%
%
%
\begin{figure}
\includegraphics[width=6.0cm,angle=270]{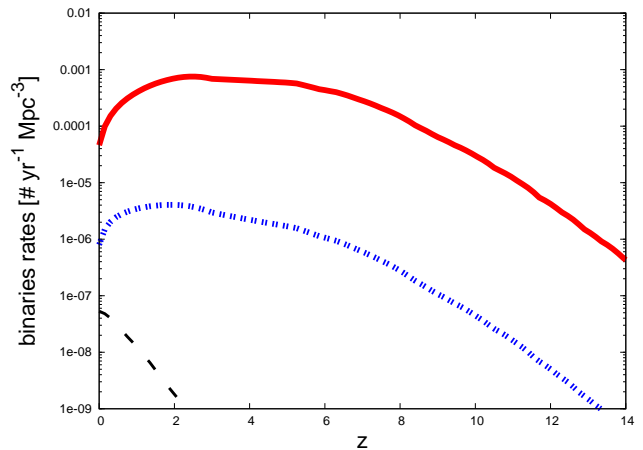}
\caption{(color online) Redshift evolution of the total binary birth rate (solid red line) and of
(BH-BH) birth and merger rates (dotted and dashed, respectively).} 
\label{binrate}
\end{figure}
Following \cite{SchFerMat2001}, we assume that a ZAMS binary forms at a
redshift $z_{s}$; after a time interval $\tau_{s}$ the system has
evolved into a (BH-BH) binary. Consequently, the redshift of formation
of the degenerate binary system, $z_{f}$ is defined through
$t(z_{f})=t(z_{s})+\tau_{s}$. Once the (BH-BH) binary system is formed,
it evolves according to gravitational wave emission until, after a time
interval $\tau_{m}$, it eventually coalesces.  The redshift $z_{c}$ at
which coalescence occurs is defined by $t(z_{c})=t(z_{f})+\tau_{m}$.
The number of (BH-BH) systems formed per unit time and comoving volume
at redshift $z_{f}$ is,
\beq \dot{R}^{\rm birth}_{\rm
(BH-BH)}(z_{f})&=&\int{d\tau_m}\int^{t(z_{f})-t(z_F)}_0{d\tau_s}\\\nonumber
&\times& \frac{N_{\rm (BH-BH)}}{N} \frac{\dot{R}_{\rm
bin}(z_{s})}{(1+z_{s})}p_{\rm (BH-BH)}(\tau_s,\tau_m), \label{BR} \eeq
where $z_{F}$  defines the onset of star formation ($z_F \sim 20$ in the
simulation), $N_{\rm (BH-BH)}$ is the number of (BH-BH) systems, $N$ is
the total number of simulated binaries, and $p_{\rm
(BH-BH)}(\tau_s,\tau_m)$ is the joint probability distribution of delay
times.  Similarly, the number of (BH-BH) systems per unit time and
comoving volume which merge at redshift $z_{c}$ is,
\beq 
\label{MR} &&\dot{R}^{\rm merger}_{\rm (BH-BH)}
(z_{c})=\int^{t(z_{c})-t(z_{F})}_{0}{d\tau_m}\int^{t(z_{c})-\tau_m-t(z_F)}_0{d\tau_s}
\nonumber\\ &&\times \frac{N_{\rm (BH-BH)}}{N} \frac{\dot{R}_{\rm
bin}(z_{s})}{(1+z_{s})}p_{\rm (BH-BH)}(\tau_s,\tau_m).  
\eeq
In Fig.~\ref{binrate} we show the redshift evolution of the binary birth
rate (solid line). Of all these systems, only 1.7 \% form a (BH-BH)
binary. The evolution of the (BH-BH) birth and merger rates is also
shown in the figure (dashed and dotted lines, respectively).  Since
$\tau_{s}$ is relatively short (less than $\sim 6$~Myr), the evolution
of $\dot{R}^{\rm birth}_{\rm (BH-BH)}$ is simply a scaled-down version
of $\dot{R}_{\rm bin}$. Conversely, there is a shift in the
evolution of $\dot{R}^{\rm merger}_{\rm (BH-BH)}$ which is due to the
long merger timescales, $\tau_m > [1-5]$~Gyr.

\begin{table} 
\begin{center} \resizebox{0.49\textwidth}{!}{
\begin{tabular}{|c|c|c|} \hline \multicolumn{3}{|c|}{Galactic
Birth/Merger Rates}\\ \hline \hline Type&Birth rates (yr$^{-1}$) &
Merger Rates (yr$^{-1}$)\\ \hline (NS-NS)& 8.2$\times 10^{-5}$
&2.0$\times 10^{-5}$ \\ \hline (BH-NS)& 5.3$\times 10^{-5}$ & 6.2$\times
10^{-6}$\\ \hline (BH-BH)& 9.5$\times 10^{-5}$ &1.8$\times 10^{-6}$\\
\hline \end{tabular}} \end{center} \caption{Galactic Birth/Merger rates
obtained from the simulation of Model A normalizing to a Galactic
supernova rate of $1\times10^{-2}$yr$^{-1}$ (see text).} \label{GBRMR}
\end{table}

For the sake of comparison, we calculate the Galactic birth and
merger rates for (BH-BH), (BH-NS) and (NS-NS) systems extracted from
the simulation of Model A. To compute these rates, we normalize
the total number of core-collapse SNe that we find in the simulation to
an estimated Galactic supernova rate of $1\times10^{-2}$yr$^{-1}$ \cite{CapEvaTur1999}.
The resulting values are presented in Table~\ref{GBRMR} and are in good agreement
with the current literature on the subject (see \cite{PorYun1998, NelYunPor2001a,
VosTau2003,RegdeF2006} and \cite{PosYun2006} and references therein). 

\section[]{GWB from black hole binaries}
\label{sec:gwback}
Following \cite{SchFerMat2001,MarSchFer2009}, we write the spectral
energy density of
the GWB produced by a population of (BH-BH) binaries as,
\be
\frac{dE}{dS df dt} = \int_{0}^{z_F} d\dot{N}^{\rm birth}_{\rm (BH-BH)}
\big{<} \frac{dE}{dS 
df} \big{>},
\ee
where $d\dot{N}^{\rm birth}_{\rm (BH-BH)} = \dot{R}^{\rm birth}_{\rm
(BH-BH)} \frac{dV}{dz} dz$ and the locally measured average GW energy
flux from a single source  at redshift $z$  is
\be
\big{<} \frac{dE}{dS df} \big{>} = \frac{c^3}{G} \frac{\pi}{2} f^2
(1+z)^2 |h[f(1+z)]|^2.
\label{eq:singlesource}
\ee

Here $f = f_e (1+z)^{-1}$ is the redshifted emission frequency 
and $h$ is the amplitude of the GW signal given in Eq.~(\ref{signal}).
The GWB is
conventionally characterized by the dimensionless quantity $\Omega_{\rm
GW}(f) \equiv {\rho_{cr}}^{-1}(d\rho_{gw}/d\log f)$, which is related to
the spectral energy density by the equation,
\be
\Omega_{\rm GW}(f)=\frac{f}{c^3\rho_{cr}}\left[\frac{dE}{dS df
dt}\right] ,
\ee
\noindent
where $\rho_{cr}=3H_0^2/8\pi G$ is the cosmic critical density.

\begin{figure}[ht]
\includegraphics[width=5.0cm,angle=270]{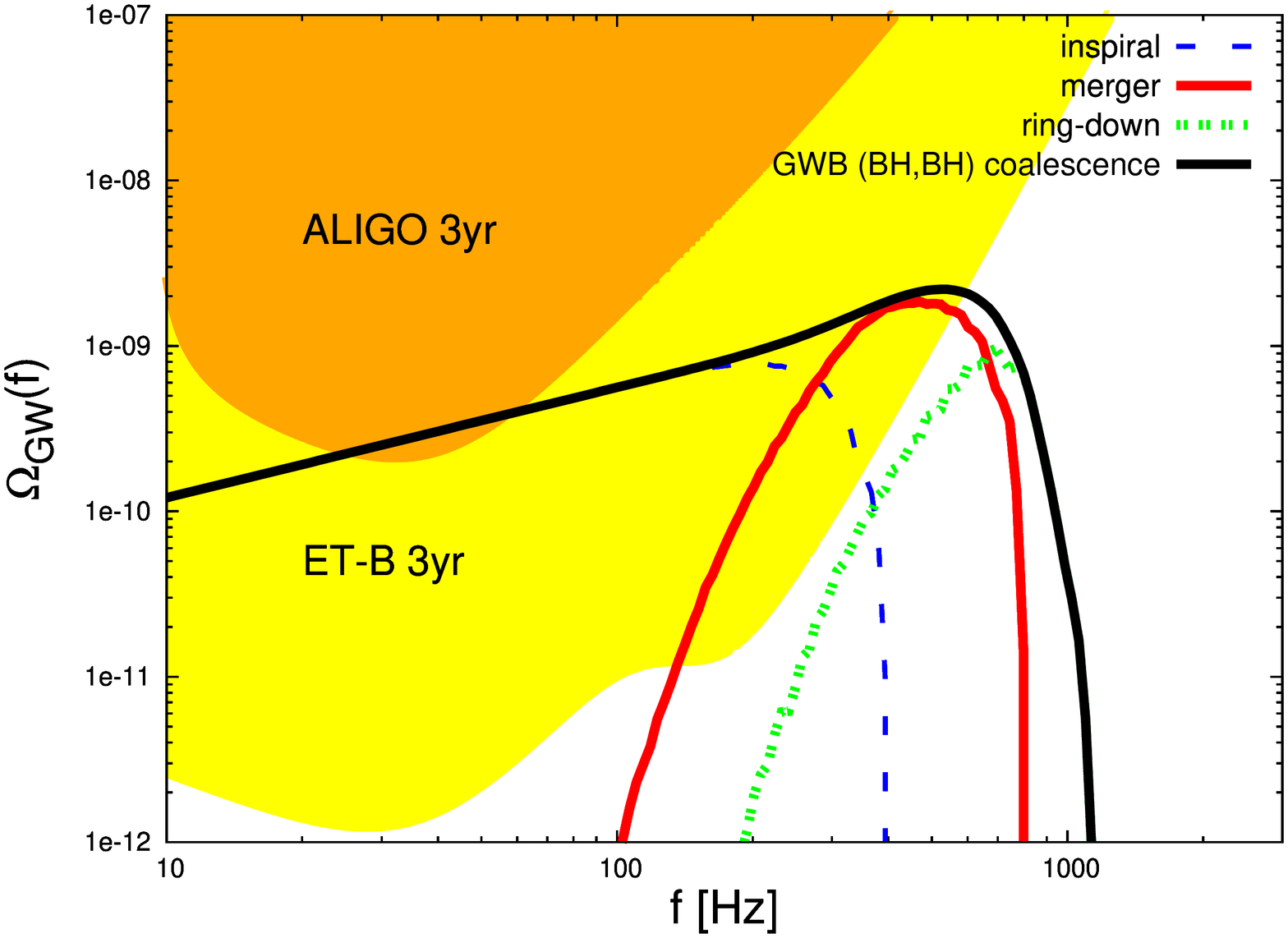}
\includegraphics[width=5.0cm,angle=270]{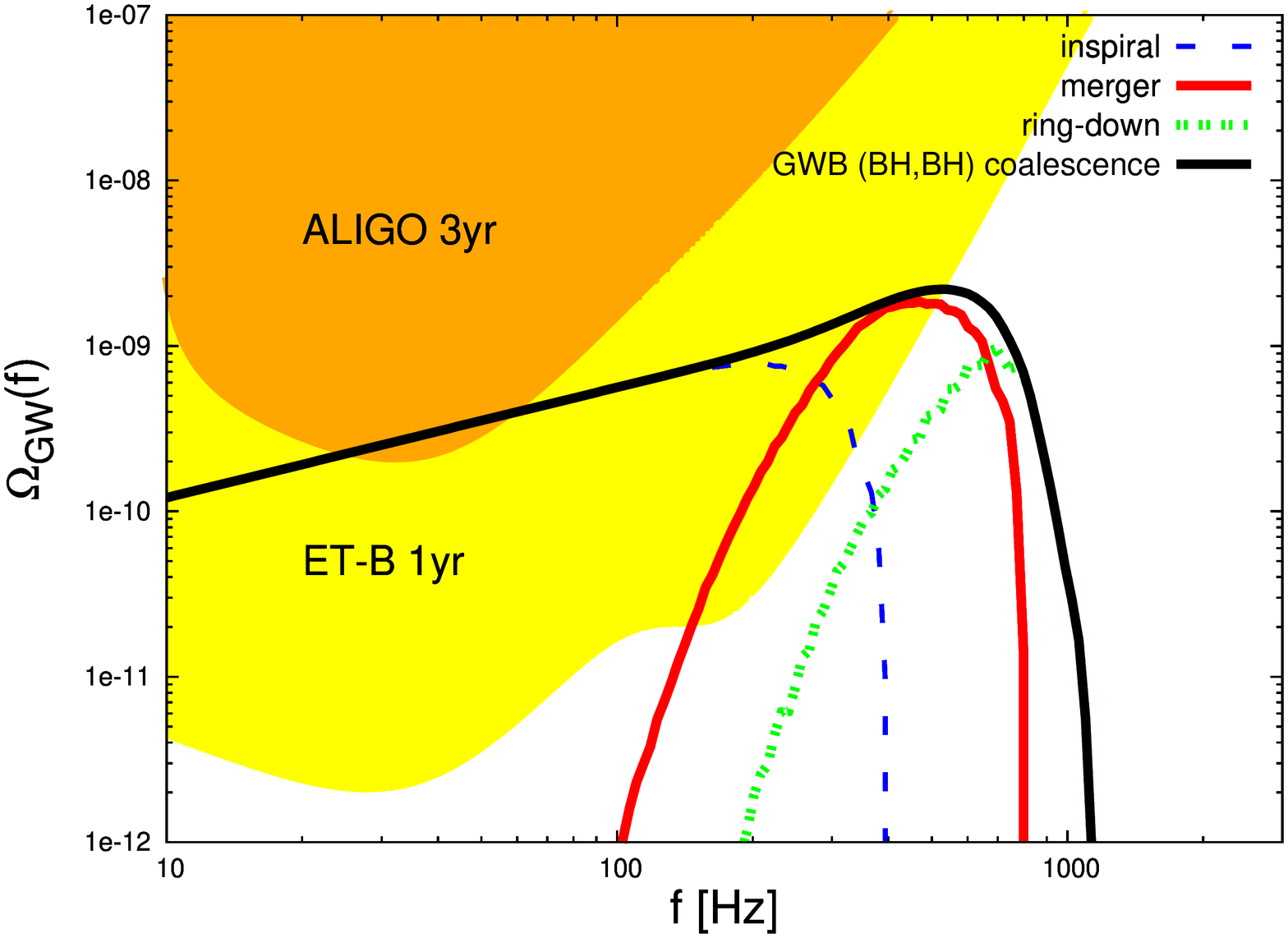}
\includegraphics[width=5.0cm,angle=270]{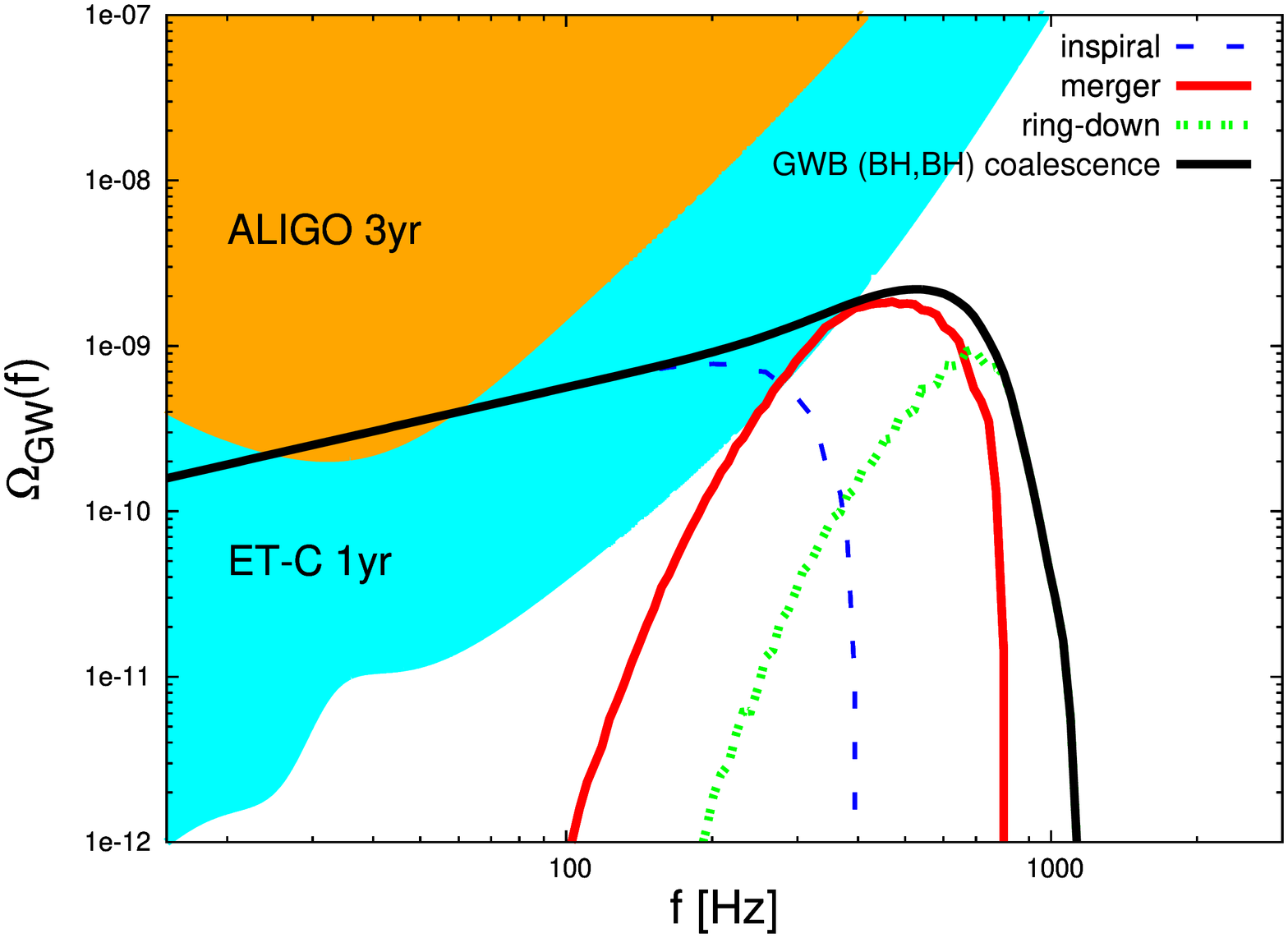}
\caption{(color online) The closure energy density,
$\Omega_{\rm GW}$, generated by
(BH-BH) coalescing binaries in model A,  plotted  as a function of the
observational
frequency.
The three contributions coming from
the inspiral (dashed), merger (solid) and ring-down (dotted)
phases are plotted separately. The black solid line is the
total GWB signal. 
The shaded regions in the three panels  indicate the foreseen
sensitivities for different interferometers and  integration times 
(see labels on the figures).
}
\label{bhbh1}
\end{figure}

In Fig.~\ref{bhbh1} we plot $\Omega_{\rm GW}$, as a function of the
observational frequency, for the reference model A.  The
cumulative signal is the result of the emission during the inspiral
(dashed line), merger (solid line) and ring-down (dotted line) phases of
the coalescence processes.  In the frequency range 10 Hz $\lesssim f
\lesssim $ 200
Hz, the signal is dominated by the inspiral phase which reaches a
maximum amplitude of $\Omega_{\rm GW}=7.8\times 10^{-10}$ at a frequency
of $\sim 200$~Hz. 
Above this limit, a further increase in the signal is
driven by the emission during the merger phase, which reaches a maximum
 $\Omega_{\rm GW} = 2.1 \times 10^{-9}$ at 540 Hz.  This is
not surprising, since a significant portion of GW energy
is radiated during the merger.  At larger frequencies, the signal
drops with a minor contribution coming from the ring-down phase which
follows the final coalescence of the two black holes in each binary.

In Fig.~\ref{bhbh1}, the GWB signal is compared with the
foreseen sensitivity curves for advanced LIGO/Virgo (ALIGO) and for the
Einstein Telescope with two different design configurations (ET-B and
ET-C) and assuming 1 yr or 3 yrs of integration time. In particular, ALIGO with 3
years of integration might sample only a small portion of the inspiral
phase with (S/N) ratios below the detection threshold (see section
\ref{sec:dect}). The full inspiral and merger phases might be observed
with ET, even with an integration time of 1 year.  The best
configuration appears to be ET-B (see discussion below) which amplifies
the sensitivity at larger frequencies.  In section \ref{sec:dect}, we
estimate the detectability of the signal in a quantitative way and in
section \ref{sec:par} we discuss the dependence of the GWB and of its
detectability on some key physical parameters.

\subsection[]{Detectability}
\label{sec:dect}
We consider the design sensitivities of second generation
interferometric detectors, Advanced LIGO/Virgo \footnote{The ALIGO
sensitivity curve is described in the public LIGO document ligo-t0900288
(https://dcc.ligo.org/public/0002 /T0900288/002/AdvLIGOnoisecurve.pdf)},
in a configuration of zero-detuning of the signal recycling mirror, with
high laser power. For the third generation interferometer ET, we
consider two target sensitivities. The first configuration, ET-B, is an
underground based design, incorporating long suspension, cryogenics, and
signal power recycling.  The second configuration, ET-C, is called {\it
Xylophone} configuration and merges the output of two detectors
specialized in different frequency bands (for more details see
\cite{HowRegCor2011,Hil2010}).

It is known that the detection strategy for continuous GWB signals is to
cross-correlate the output of two detectors that are assumed to be
sufficiently well separated that their noise sources are largely
uncorrelated \footnote{For ET this condition is not satisfied, but new
techniques to remove instrumental correlation are under development
\cite{HowRegCor2011}.}.

The statistical nature of the background depends on the duration of the
signal, on the event rate and on the lower frequency bound of the
detector, $f_{\rm L}$, the so-called "seismic wall" (see for a
discussion \cite{RegHug2009}).  It has recently been suggested  that,
whenever the signal is very short (as it is expected for binaries at
high frequencies), and of sufficiently high amplitude,
 the resolution of the detector or of the data
analysis method will enable to individually identify the signals and
subtract it from the data \cite{Ros2011}.

Assuming that for ALIGO $f_{\rm L} = 10$~Hz, the background predicted
for the reference model A would be characterized by a duty cycle of
$\sim 5 \times 10^{-3}$. For ET, which has a lower $f_{\rm L} = 1~$Hz,
the duty cycle would be  $\sim 2$.  According to \cite{DraFla2003}, the
cross-correlation method is found to be nearly optimal  for duty cycles
$> 10^{-3}$. Therefore, in what follows we estimate the detectability of
the signal using this method, as in \cite{ZhuHowReg2011}, and we do not
distinguish between the resolvable and the non-resolvable components
\cite{Ros2011}.
\begin{table}[ht]
\begin{tabular}{|c|c|c|c|c|}
\hline
\multicolumn{5}{|c|}{$(S/N)$}\\
\hline
\hline
\multicolumn{5}{|c|}{Case-II [Case-I]}\\
\hline
GWB&ALIGO & ET-B  & ET-B &ET-C \\
& (3 yr)&  (3 yr) &  (1 yr)& (1 yr)\\
\hline
coalescence&1.9[0.7]&316[118]&182[68]&275[103]\\
\hline
inspiral&1.9[0.7]&310[116]&179[67]&274[103]\\
\hline
merger&7.5$\times 10^{-2}$[6.4$\times 10^{-4}$]&26[9]&15[5.6]&5[1.9]\\
\hline
ring-down&8.1$\times 10^{-3}$[3.3$\times
10^{-5}$]&1.5[0.6]&0.9[0.3]&0.4[0.1]\\
\hline
\end{tabular}
\caption{The $(S/N)$ ratio for second and third generation detectors,
assuming different integration times and detector separation/orientation
(see text).  The values refer to model A (see table \ref{ModelA})
and have been computed considering the cumulative signal
(coalescence) and separate contributions from the inspiral, merger and
ring-down phases.}
\label{tableSNR}
\end{table}
We use the cross correlation statistics to calculate the optimized 
$S/N$ for an integration time $T$ as given by \cite{AllRom1999},
\be
\left(\frac{S}{N}\right)^2 \approx\frac{9H^4_0}{50\pi^4}T\int^{\infty}_0 df 
\frac{\gamma^2(f)\Omega^2_{\rm GW}(f)}{f^6P_1(f)P_2(f)} \, ,
\label{snr}
\ee
\noindent
where $P_{1}(f)$ and $P_{2}(f)$ are the power spectral noise 
densities of the two detectors and $\gamma$ is the normalized 
overlap reduction function, which quantifies the loss of sensitivity 
due to the separation and the relative orientation of the detectors. 

We have computed the $(S/N)$ ratio assuming different integration
times (1 - 3 years) and detector separation/orientation (case-I and -II). 
For ALIGO, case-I considers the LIGO Hanford/Livingston pair using
$\gamma$ in the form given by Eq. (3.26) in \cite{AllCreFla2002}. 
For ET-B and ET-C, case-I adopts the constant value $\gamma = -3/8$
which applies to two ET detectors  operating
in the frequency range [1-1000]~Hz \cite{HowRegCor2011}. Case-II
is the same for all detectors and represents a pair of aligned 
equivalent detectors situated within several km. This optimal case
corresponds to $\gamma = 1$. 

\begin{table*}
\begin{center}
\begin{tabular}{|c|c|c|c|c|}
\hline
\multicolumn{5}{|c|}{EXPLORATION OF THE PARAMETER SPACE}\\
\hline
\hline
 Model & Modified Parameter & Galactic (BH-BH) & Galactic (BH-BH) & Local (BH-BH)\\
 &     & BR (Myr$^{-1}$)&MR (Myr$^{-1}$)&MR (Mpc$^{-3}$Myr$^{-1}$)\\
\hline
 A  & reference model& $95$&$1.8$& $5.3\times10^{-2}$\\
\hline
 B1 &$\alpha_{CE}\lambda$=0.5&$95$&$16$& $2.9\times10^{-1}$ \\
\hline
 B5 &$\alpha_{CE}\lambda$=4&  $96$& $1.2\times10^{-1}$ & $3.3\times10^{-3}$ \\
\hline
 C1 & Paczy\`nsky distribution ($\sigma=150$ km $s^{-1}$)& $120$& $1.7$&$5.1\times10^{-2}$ \\
\hline
 C2 & Paczy\`nsky distribution ($\sigma=600$ km $s^{-1}$)& $71$&$2.4$&$6.6\times10^{-2}$\\
\hline
 C5 & Maxwellian distribution ($\sigma=200$ km $s^{-1}$)& $88$& $1.8$&$5.3\times10^{-2}$\\
\hline
 C7 & Maxwellian distribution ($\sigma=600$ km $s^{-1}$)& $53$& $3.2$& $8.8\times10^{-2}$\\
\hline
 C8 & No kick distribution& $220$& $1.7$&$5.3\times10^{-2}$\\
\hline
 F1 & $m_{\rm thre,BH}=8.5 M_{\odot}$ & $140$& $7.6$&$2.0\times10^{-1}$\\
\hline
 F2 &  $m_{\rm thre,BH} = 7.6 M_{\odot}$ & $160$ &$15$&$3.6\times10^{-1}$\\
\hline
 F3 & $m_{\rm thre,BH} = 5.5 M_{\odot}$ & $240$& $40$&$8.5\times10^{-1}$\\
\hline
\end{tabular}
\caption{
In columns 3 and 4 we  give the  Galactic Birth (BR)/Merger rates (MR) 
for the different models, obtained by varying one key parameter (indicated in
colum 2) with respect to the reference model A (see text).
In column 5 we give the local merger rate, which is the $z=0$ value of
the merger rate (see for instance Fig. \ref{binrate} for model A)
obtained as described in Section \ref{sec:rates}, Eq. (\ref{MR}).
}
\label{diffModels}
\end{center}
\end{table*}
The resulting $(S/N)$ ratios are reported in Table~\ref{tableSNR}.
The values  refer to model A (see table \ref{ModelA}),
and are obtained assuming a threshold signal-to-noise ratio of 3
which corresponds to a false alarm rate of 10\% and to a detection rate of
90\% (for more details see eq. 19 in \cite{MarCioSch2011}).

The highest $(S/N)$ ratios are obtained with
optimal orientation (case-II) and longer integration times.
These conditions would not allow ALIGO to detect the coalescence signal.
The increase in sensitivity foreseen for ET would enable the
detectability of two portions
of the signal, the inspiral and the merger phases, with $(S/N) >
5$, independently of
the adopted separation/orientation and integration times.
\subsection[]{Dependence on physical parameters}
\label{sec:par}

In this section we analyze the dependence of the GWB on some key physical 
parameters that affect the (BH-BH) birth/merger rates, i.e. the adopted 
kick velocity distribution, the CE parameter and core mass threshold for
black hole formation, $m_{\rm thre,BH}$. 

In Table \ref{diffModels} we list a set of models (first column) which
differ from the reference model A by the variation of a single
parameter, indicated in the second column.  For each model, in column 3
and 4  we tabulate  the Galactic (BH-BH) birth and merger rates.  
Similarly to what done in Section \ref{sec:rates} (Table \ref{GBRMR}),
these
values are obtained by normalizing the  supernova rate predicted by each
SeBa run  to the Galactic
rate of $1\times 10^{-2}$ yr$^{-1}$.  
In column 5 we give the local merger rate, which is the $z=0$ value of
the cosmic merger rate (see for instance Fig. \ref{binrate} for model A)
obtained as described in Section \ref{sec:rates}, Eq. (\ref{MR}).

In Table 8 of ref.\cite{AbaAbbAbb2010}, the Galactic rates obtained
with different population synthesis codes, based on different assumptions
on stellar/binary evolution, are compared. The values reported in that
table range between $0.01$ and $250$  Myr$^{-1}$, and 
values that are considered as ``realistic'' range within $0.01$ and
$~20$  Myr$^{-1}$. If we  compare 
these estimates with the Galactic merger rates  given in column 4
of Table \ref{diffModels}, we see that our values are in the realistic
range, except for the model F3, which we will discuss below.

We first consider  models where we vary the common envelope parameter (models
B1 and B5).  This parameter does not affect the birth rate but controls
the number of merging systems: in fact, a larger (smaller) CE parameter,
such as in model B5 (B1), generates binaries which, at the end of the CE
phase, are characterized by larger (smaller) orbital separations and
therefore longer (shorter) merger timescales.  The resulting fraction
of merging systems in model B5 is only 0.12\% of the total (BH-BH)
binaries, more than a factor 10 smaller then for the reference model A.  

The adopted shape and velocity dispersion of the kick distribution
affect both the birth and merger rates (see models C1, C2, C5 and C7).
In the first two models, we assume a Paczy\'nsky distribution for kick
velocities, \be P(u)du=\frac{4}{\pi}\frac{du}{(1+u^2)^2} \label{Pac} \ee
where $u= v/\sigma$ and the dispersion velocity $\sigma$ is,
respectively, 150 and 600 km s$^{-1}$ for models C1 and C2. In models C5
and C7 we consider a Maxwellian distribution, \be P(v)dv=
\sqrt{\frac{2}{\pi}}v^2 e^{-\frac{v^2}{2\sigma^2}} \label{Max} \ee with
$\sigma= 200$ and $600$ km s$^{-1}$, respectively.
For both distributions, the birth rate decreases with increasing
$\sigma$.  However, the fraction of formed binaries which coalesces
increases slightly with $\sigma$.  The first effect is a direct
consequence of the disruption of the binary after the SN explosion (it
is more likely that a strong kick disrupts the system).  The increase of
the merger rate with $\sigma$ is due to the net effect of kicks on the
orbit of systems which are not disrupted: while the semi-major axis is
left mostly unchanged, the eccentricity is greatly enhanced, favouring
the orbital decay due to GW emission.  As a result, in model C2 the
number of merging pairs is a factor 2 larger than in model A (3.4\% of
the total sample).  In the extreme case of null kicks, such as in model
C8, we find a birth rate which is a factor 2.3 larger than in model A.
Yet, the corresponding merger rates are comparable in the two models.

Finally, we discuss the dependence on the core mass
threshold for BH formation (models F1-F3).  This parameter appears to be
the most important one for the GWB.  In fact, a reduction in $m_{\rm
thre,BH}$ from the reference value of $10 M_{\odot}$ (model A) to $5.5
M_{\odot}$ (model F3) leads to a birth rate which is a factor 2.5
larger. The amplification in the merger rate is even more dramatic, by
more than a factor 20.  This is due to the evolutionary path followed by
massive stellar progenitors.  In particular, a smaller $m_{\rm thre,
BH}$ allows the formation of BHs from lighter progenitor stars;
the latter experience a smaller amount of mass loss (see Section
\ref{sec:seba}) forming close binary pairs
characterized by shorter merger timescales.

\begin{figure}[ht]
\includegraphics[width=6.0cm,angle=270]{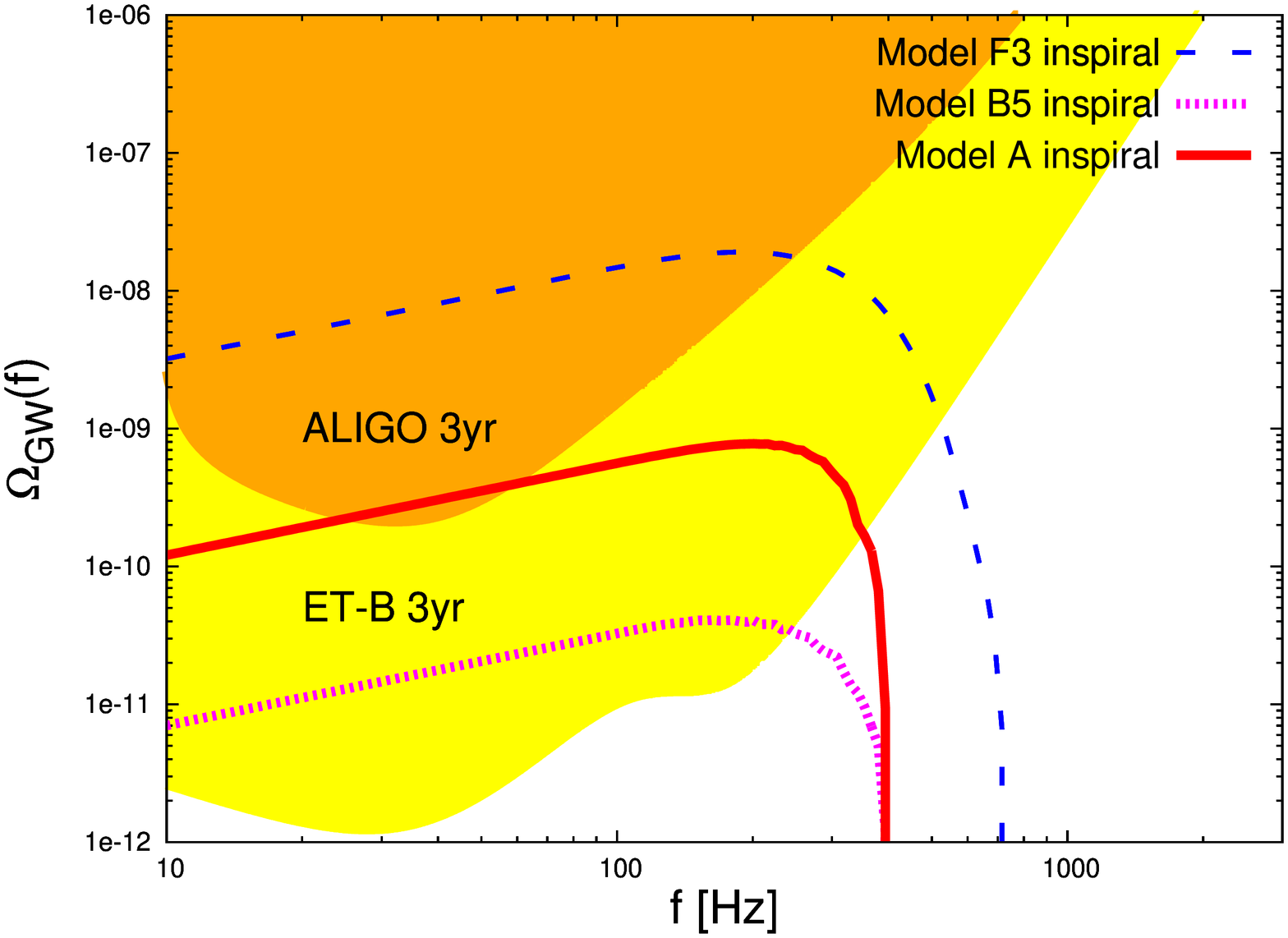}\\
\includegraphics[width=6.0cm,angle=270]{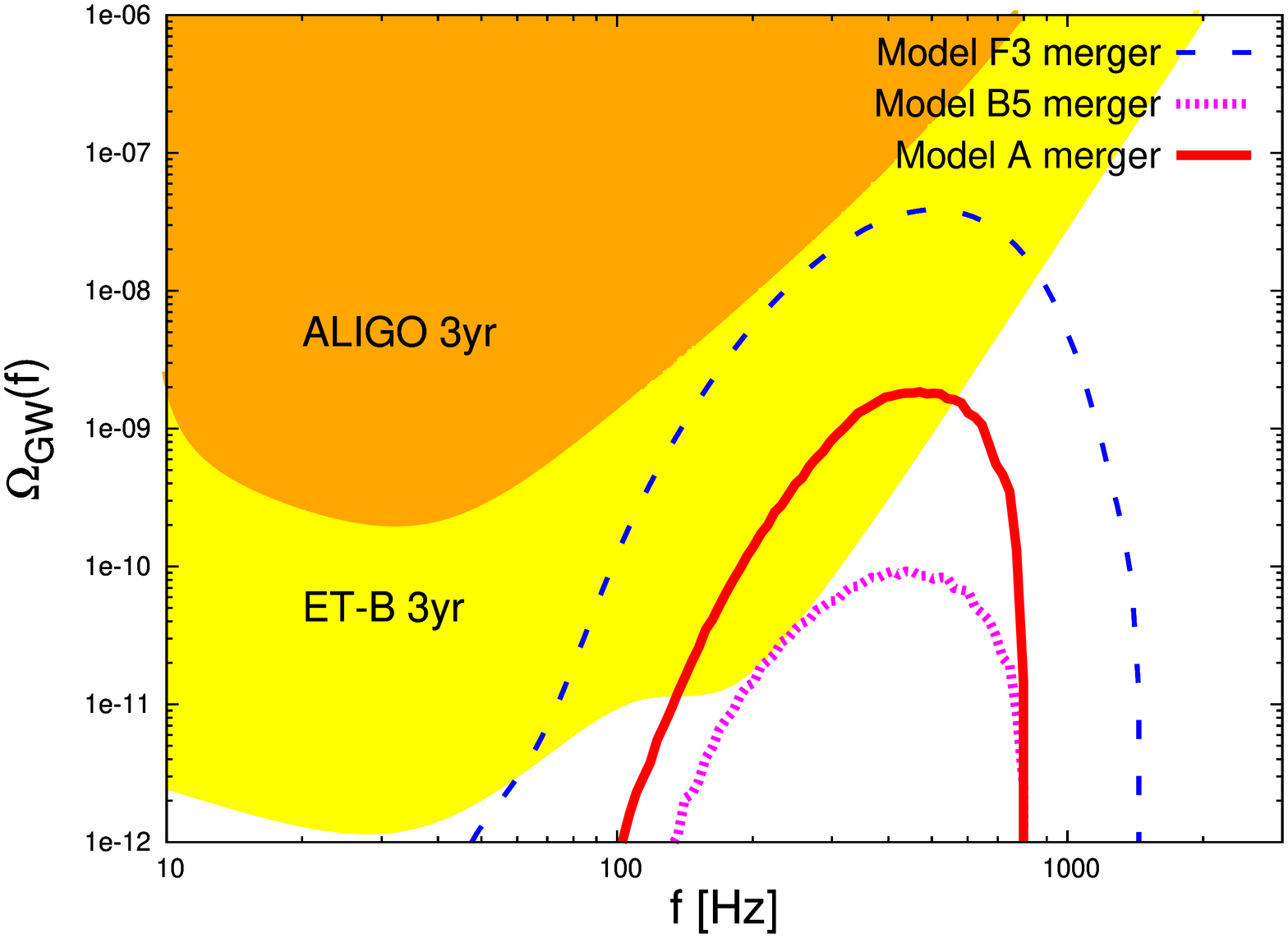}\\
\includegraphics[width=6.0cm,angle=270]{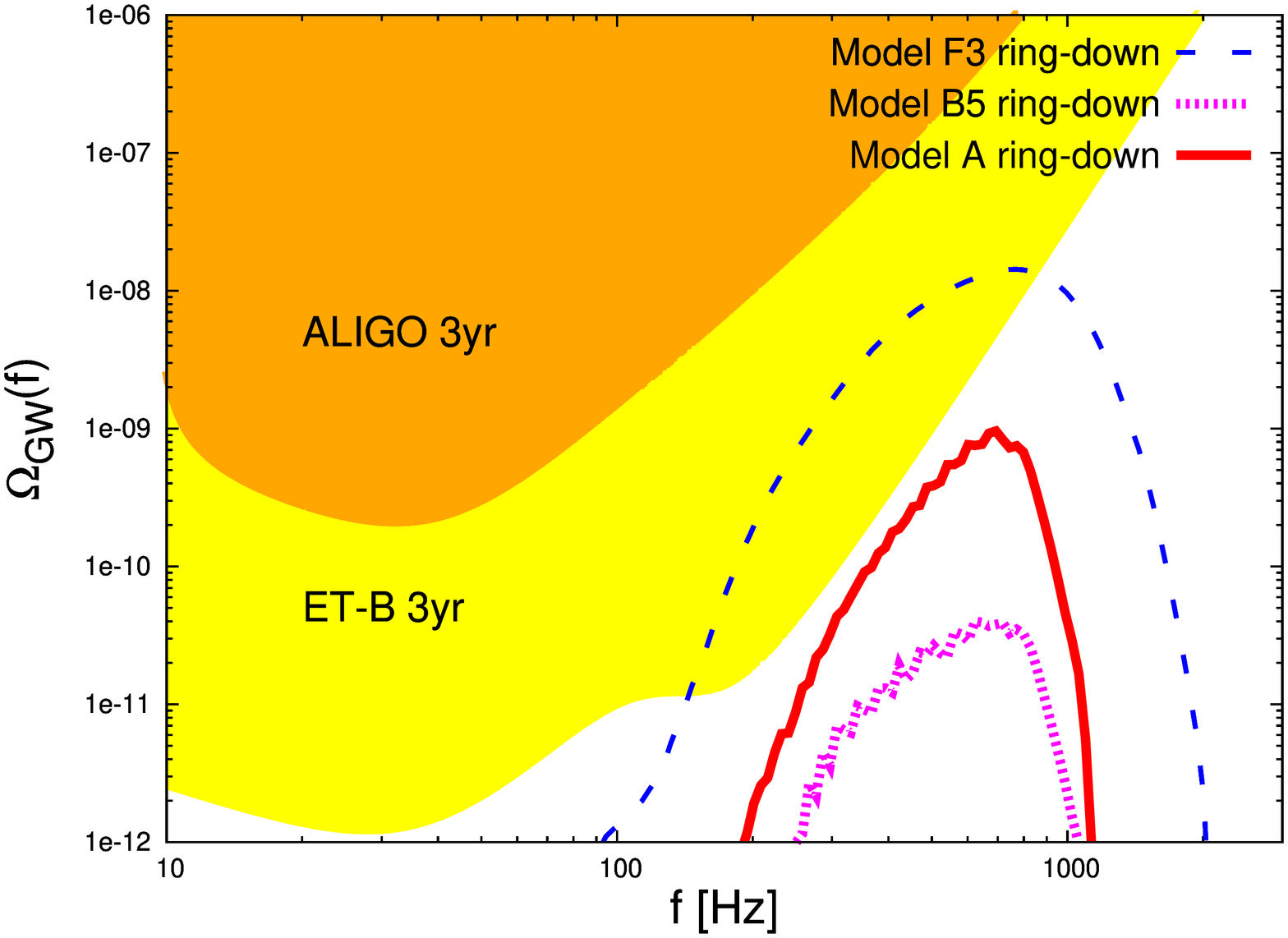}
\caption{(color online)$\Omega_{\rm GW}$ generated during the
inspiral ({\it upper} panel), merger ({\it central} panel) and ring-down
({\it lower} panel) phases
in models F3 (dashed), B5 (dotted) and A (solid).
In all panels, the two shaded regions indicate the foreseen
sensitivities
of ALIGO and ET-B assuming 3 years of integration.}
\label{confF3B5}
\end{figure}

\begin{figure}
\includegraphics[width=6.0cm,angle=270]{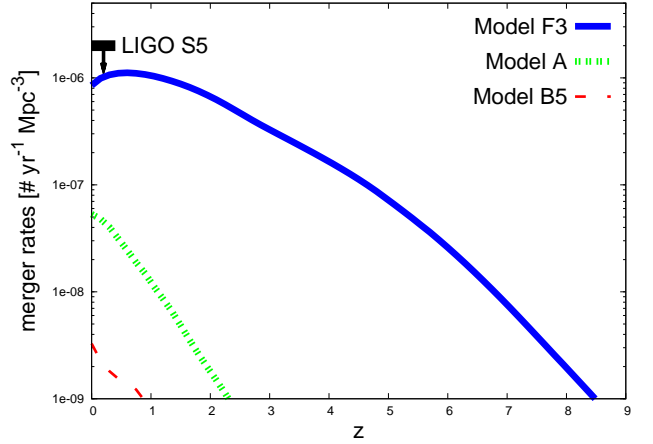}
\caption{(color online) Redshift evolution of
(BH-BH) merger rates for models F3 (solid line),  A (dotted line) 
and B5 (dashed line). The upper limit labeled `LIGO S5' shows the constraint
recently derived in \cite{Abadieetal2011} (see text).} 
\label{mergconf}
\end{figure}
The largest differences with respect to model A are found for models B5
and F3 which provide a sort of lower and upper limits to the (BH-BH)
merger rate and GWB. In Fig.~\ref{mergconf} we show the predicted
redshift evolution of (BH-BH) merger rate in the two models as compared
to model A.  The upper limit labelled as `LIGO S5' shows the constraint derived in
\cite{Abadieetal2011} from approximately 2 years of LIGO data (run S5) on the
merger rate of systems with component masses in the range $19 M_{\odot}
- 28 M_{\odot}$.  Even the most optimistic model F3 predicts a local
merger rate which is more than a factor 2 smaller than the 
upper limit  inferred by the data analysis of the LIGO S5 run.

\begin{figure}
\includegraphics[width=6.0cm,angle=270]{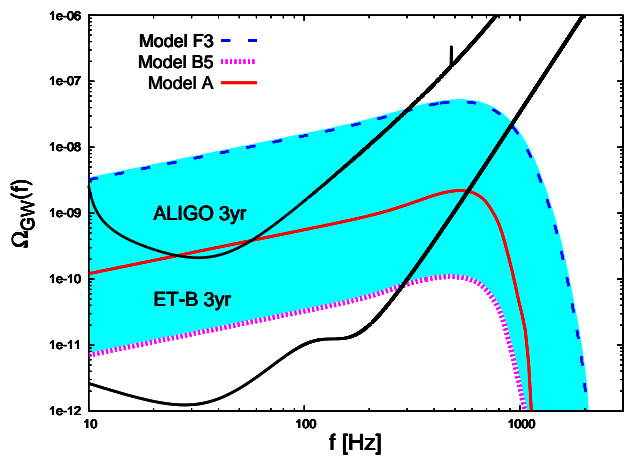}
\caption{(color online) The closure energy density, $\Omega_{\rm GW}$, for 
(BH-BH) binaries predicted by models F3 (dashed line) and B5 (dotted line) 
is compared to that of model A (solid line). The
shaded region can be considered  a measure of the uncertainty on $\Omega_{\rm GW}$.
The two black solid lines indicate the foreseen sensitivities of ALIGO and ET-B assuming 
3 years of integration.}
\label{GWBband}
\end{figure}

The longer (shorter) merger timescales predicted in model B5 (F3) lead
to a reduction (amplification) of the overall cosmic merger rate,
shifting it to smaller (larger) redshifts.  This, in turn, affects the
amplitude and frequency range of the resulting GWB spectra, as shown in
Figs.~\ref{confF3B5} and \ref{GWBband}.    

In Fig.~\ref{confF3B5}, we plot the contributions to $\Omega_{\rm
GW}$ generated  during the inspiral (upper panel), merger (central panel)
and ring-down (lower panel) phases comparing models F3, B5 and A. 
As expected, model F3 generates the strongest signals.  It is also evident
that the merger and ring-down signals in model F3 extend to lower
frequencies with respect to models A and B5.  This is due to the shorter
merger timescales which allow a larger number of (BH-BH) binaries to
reach the final coalescence at larger redshifts (see
Fig.~\ref{mergconf}), emitting signals contributing  at smaller
observational frequencies.  Similarly, the differences among models A
and B5 can be traced back to the merger timescales which, for models B5,
confines (BH-BH) coalescence to redshifts $z < 1$.  It is also
interesting to note that while models A and B5 show a similar behaviour
at the largest frequencies, model F3 systematically extends to larger
frequencies. In fact, the smaller black hole masses predicted in model
F3 (as a consequence of the smaller $m_{\rm thre, BH}$) lead to larger
$f_{\rm merge}$, $f_{\rm ring}$, and $f_{\rm cut}$.

The shaded region in Fig.~\ref{GWBband} illustrates the largest
variations of the GWB among 
the models,  and can be viewed as an indication of 
the uncertainty affecting its estimate. For the models we consider,
the peak amplitude in the closure
energy density ranges within $10^{-10} \leq \Omega_{\rm GW} \leq 5 \times 10^{-8}$
at frequencies $470$~Hz $\leq f \leq 510$~Hz.

Table~\ref{tableSNR2} quantifies these differences in terms of the
predicted signal-to-noise ratio. For the sake of comparison, we consider
the same integration times and detector configurations/orientations as
in Table~\ref{tableSNR}, which refer to model A. The major difference
with respect to the data in Table~\ref{tableSNR} is that, for model F3,
ALIGO has a chance to probe the inspiral part of the GWB with an
integration time of 1 yr and $(S/N)$ larger than 10.

\begin{table*}
\begin{center}
\begin{tabular}{|c|c|c|c|c|c|}
\hline
\multicolumn{6}{|c|}{$(S/N)$}\\
\hline
\multicolumn{6}{|c|}{Case-II [Case-I]}\\
\hline
GWB& Model &ALIGO (1 yr)&ALIGO (3 yr)& ET-B (1 yr) & ET-B (3 yr)\\
\hline
inspiral& F3 &29[10]&50[19]&4720[1770]&8176[3065]\\
\hline
inspiral& B5 &6.4$\times 10^{-2}$[2.4$\times 10^{-2}$] &0.1[4.1$\times 10^{-2}$]&10[3.9]&18[6.7]\\
\hline
merger&F3 &1.1[1.3$\times 10^{-2}$]&1.9[2.3$\times 10^{-2}$]&442[165]&766[287]\\
\hline
merger&B5 &3.0$\times 10^{-3}$[3.3$\times 10^{-5}$]&5.2$\times 10^{-3}$[5.6$\times 10^{-5}$]&1.2[0.4]&2.1[0.8]\\
\hline
ring-down&F3&0.1[7.0$\times 10^{-4}$]&0.2[1.2$\times 10^{-3}$]&32[12]&56[21]\\
\hline
ring-down&B5&3.3$\times 10^{-4}$[1.7$\times 10^{-6}$]&5.7$\times 10^{-4}$[2.9$\times 10^{-6}$]&8.2$\times 10^{-2}$[3.0$\times 10^{-2}$]&0.1[5.3$\times 10^{-2}$]\\
\hline
\end{tabular}
\end{center}
\caption{The $(S/N)$ ratio for second and third generation detectors assuming
different integration times and detector separation/orientation.
These values 
 have been computed considering separate contributions from the inspiral, 
merger and ring-down phases for models F3 and B5.
The numbers in square brackets refer to a non optimal
orientation of the detectors (see the discussion in
Sec.~\ref{sec:dect}).
}
\label{tableSNR2}
\end{table*}
\section{Discussion and conclusions}
\label{sec:conclusions}
Many astrophysical processes which control the formation and evolution of (BH-BH)
binaries, starting from their stellar progenitors, are still poorly
understood. Of particular relevance in this respect, 
are the amount of mass loss by massive stars,
pre-supernova and supernova evolution, the effects of mass transfer
among the two companion stars on the subsequent evolution of the system.
In the present study, we have considered a reference model (A), which 
adopts a set of ``standard'' conservative assumptions, reproducing 
the observed properties of single Wolf-Rayet stars and double pulsars
(\cite{PosYun2006} and references therein). 
Furthermore, we have explored a wide range of parameters on which our
simulations depend (see Table 1), to extract those that have the
largest impact on the GWB, i.e. the common envelope parameter, the
core mass threshold for BH formation and the kick velocity distribution.
Varying these parameters, we identify two models, B5 and F3, which
produce, respectively, the smallest and largest gravitational wave
background.

When normalized to a Galactic star
formation rate, the  Galactic birth and merger rates computed for the
considered models  are  in good agreement
with the results of independent studies (see \cite{AbaAbbAbb2010} and
references therein, and the discussion in Sec. \ref{sec:rates}).


To model the single source emission, we use the hybrid waveforms given
in \cite{AjiBabChe2008}, which refer to non-spinning (BH-BH) binaries.
This model has been improved in \cite{AjiHanHus2011} with the inclusion
of the effect of non-precessing spins, and of a more accurate modeling of
the non-spinning case. A comparison of the GWB obtained with the
waveforms of \cite{AjiBabChe2008} and of \cite{AjiHanHus2011}, for our
reference model A, assuming the
single spin parameter $\chi=0$ (non spinning case),
shows no significant differences. Even
assuming that all (BH-BH) binaries have $\chi=0.85$ (which is the
extreme case considered  in \cite{AjiHanHus2011}), the GWB shows some
difference only above $\sim 1$ kHz, in a region where even ET could not detect
it. This evidence has been reported also in \cite{ZhuHowReg2011}. In
\cite{AjiHanHus2011} it is stated that the non-precessing waveforms are
effectual in capturing also precessing binaries, expecially in the
comparable-mass regime. Since most of our systems have $q<1.1$ (see
figure \ref{fig:ecc}) the inclusion of spin precession is not expected
to significantly change  our results.

As mentioned in the introduction, the same waveforms have been
recently used by \cite{ZhuHowReg2011} to estimate the gravitational wave
background. This study adopts a fixed merger time distribution function,
an average chirp mass to describe all the (BH-BH) systems, and normalize
the binary merger rate to a ``local'' merger rate (LMR); the latter
quantity is computed multiplying the Galactic merger rates predicted by
Population Synthesis Models, \cite{BelDomBul2010}, by the average number
density of Milky Way-type galaxies in the Local Universe (assumed to be
0.01 Mpc$^{-3}$).  Thus, there are many differences between this study
and the present one.  As explained in section \ref{sec:rates},  we do
not normalize the cosmic merger rate to a ``re-scaled'' Galactic rate;
indeed,  we compute it integrating the star formation history over the
birth and merger time distribution functions predicted by each SeBa run.
The resulting local rates, at $z=0$, are shown in column 5 of Table
\ref{diffModels}, and are a factor 2 - 3 (depending on the model) larger
than what would be obtained from the corresponding Galactic merger rate
(column 4), applying the procedure of \cite{ZhuHowReg2011}.  In fact,
due to their long merger times, a large fraction of the (BH-BH)
progenitors form at $z \ge 1$ close to the peak of the cosmic star
formation rate. In addition, our analysis shows that there exist
correlations between the distributions of merger times, hence the merger
rate, and the distribution of the chirp masses (see
Fig.~\ref{fig:times}).  Thus, the points of the detectable parameter
space explored by \cite{ZhuHowReg2011}, identified by an average chirp
mass and a LMR, do not have the same probability to represent a physical
model. Our study shows that variations of key physical parameters
produce {\it correlated} effects on the merger time and chirp mass
distributions and, as a consequence, on the amplitude and the spectral
energy distribution of the gravitational wave background (see, for
instance, the discussion on  model F3 in section \ref{sec:par}).  These
properties can be appreciated only using the full rich information
provided by Population Synthesis Models.

Our main results can be summarized as follows:
\begin{itemize}
\item For the reference model A,
the sample of simulated (BH-BH) binaries is characterized by BH
masses which vary between $\sim 6 M_{\odot}$ and $\sim 20 M_\odot$, with
the largest concentration in the range $[10 - 15] M_{\odot}$.   The
formation of (BH-BH) binaries from their stellar progenitors is
characterized by relatively short timescales, $\sim 3.5 - 6$ Myr and by
a wide interval of semi-major axes ranging between $\sim 10 R_\odot$ to
several thousands of $R_\odot$.
Only 2\% of the formed (BH-BH) binaries
are able to merge within the Hubble time. These systems are
characterized by semi-major axes $< 20 R_{\odot}$, and black hole mass
ratios close to 1.  The majority of these systems (70\%) have merger
timescales $\geq 6$ Gyr.
As a result, (BH-BH) Galactic birth and merger rates are, respectively,
$9.5\times 10^{-5}$ yr$^{-1}$ and $1.8\times 10^{-6}$ yr$^{-1}$.
On cosmic scales, the (BH-BH) birth rate closely follows the
shape of the cosmic star formation rate (although with a significantly
reduced amplitude); conversely, the merger rate shows a significant time
delay, and it is negligible beyond $z \sim 2$. 
The above conclusions mostly depend  on the adopted common envelope
parameter and core mass threshold for BH formation ($\alpha_{CE} \lambda
= 2$ and $m_{\rm thre, BH} = 10 M_{\odot}$ in  model A).

\item
An increase of the CE parameter to $\alpha_{CE} \lambda = 4$, as in
model B5,  generates (BH-BH) binaries with larger orbital separation,
reducing the (BH-BH) merger rate (by a factor of 10 for the Galactic
value) and confining the mergers to occur at $z < 1$. Conversely, a
reduction in $m_{\rm thre, BH}$ to $5.5 M_{\odot}$, as in model F3,
leads to an increase of the Galactic merger rate by more than a factor 20. 
Variations in these physical parameters also affect the distribution
of black hole masses. 

\item  The GWB is characterized by a peak amplitude in the range 
$10^{-10} \leq \Omega_{\rm GW} \leq 5 \times 10^{-8}$ at frequencies $470$~Hz $\leq f \leq 510$~Hz,
when the uncertainties on some key physical parameters are considered 
(see Fig~\ref{GWBband}). 

\item 
Advanced LIGO/Virgo have a chance to detect the GWB from the inspiral
only in model F3, which predicts the highest merger rate; 
third generation detectors like ET, would detect the inspiral GWB with
high $(S/N)$ for models spanning the region from model A to model F3 in
Fig.~\ref{GWBband} (see Tables~\ref{tableSNR} and \ref{tableSNR2}).

\item The merger  contribution to the
GWB could be detected only by ET. From Fig.~\ref{confF3B5} and 
Tables~\ref{tableSNR} and \ref{tableSNR2}, we see that ET-B could detect this
contribution for models spanning the region from model A 
($(S/N)\gtrsim ~ 6$) to model F3 ($(S/N)\gtrsim ~ 165$), with 1 yr integration.

\item The ring-down contribution could be detected by ET-B with 1 yr
integration only for model F3 ($(S/N)\geq ~ 12$).
\end{itemize}

We find that the  amplitude of the GWB is very
sensitive to the adopted core mass threshold for BH formation (models F1
to F3). This
opens up the possibility to constrain the uncertain physics related to
the final stages of the evolution of massive stars using observational
constraints on the associated gravitational wave emission.

Finally, we would like to mention that, according to a recent study \cite{BelDomBul2010}, 
(BH-BH) binaries formed in a
low-metallicity environment are characterized by higher coalescence rates
and chirp masses than their solar metallicity counterparts.
Indeed, at sub-solar metallicities
stars are more compact (smaller radii), experience reduced mass loss
(larger masses) and BHs can form by direct collapse of the progenitor
(no kick due to the SN explosion). 
These changes result in an increase of the galactic and local merger 
rate. Metallicity-dependent evolutionary tracks are   
currently being implemented in SeBa \cite{Toonenetal2011}, and their effects
on the gravitational wave background will be considered in a future study. 

\section*{Acknowledgments}
We thank Tania Regimbau, Xing-Jiang Zhu and Pablo Rosado 
for fruitful comments which  enabled us to improve this paper.
We also acknowledge
Francesco Pannarale for his careful reading of the manuscript. 
Stefania Marassi thanks Francesco Simula for technical support.
This work was partly supported by the Netherlands Research Council NWO (via
grants \#643.200.503, \#639.073.803 and \#614.061.608) and the
Netherlands Research School for Astronomy (NOVA).

\label{lastpage}
\end{document}